\documentclass[a4paper,11pt]{article}

\usepackage[margin=2.5cm]{geometry}
\usepackage{mathptmx}
\usepackage{amssymb}
\usepackage{hyperref}
\hypersetup{bookmarksnumbered=true,
	pdftitle={A construction of the genetic material},
	pdfauthor={J.-L. Sikorav, A. Braslau, and A. Goldar},
	pdfkeywords={DNA, RNA, proteinogenic amino acids},
	}
\usepackage[nosort]{cite}
\usepackage{mdwlist}
\usepackage{multicol}
\usepackage{graphicx}
\usepackage{pdflscape}
\usepackage[german,french,italian,polutonikogreek,english]{babel}

\begin{document}

\noindent
{\Huge\bf A construction of the genetic material and of proteins}
\\[1em]
\rule{\textwidth}{1bp}\\[5bp]
{\large J.-L. Sikorav$^1$, A. Braslau$^2$ and A. Goldar$^3$}\\
\rule{\textwidth}{1bp}
\\[1em]
{\it
$^1$DSM, Institut de Physique Th{\'e}orique, IPhT, CNRS, MPPU, URA2306, CEA/Saclay, F-91191 Gif-sur-Yvette, France.
\\
$^2$DSM, Service de Physique de l'{\'E}tat Condens{\'e}, SPEC, CNRS URA2464, CEA/Saclay, F-91191 Gif-sur-Yvette, France.
\\
$^3$DSV, iBiTec-S, Service de Biologie Int{\'e}grative et de G{\'e}n{\'e}tique Mol{\'e}culaire, CEA/Saclay, F-91191 Gif-sur-Yvette, France.
}

\begin{abstract}

A theoretical construction of the genetic material establishes the unique and ideal character of DNA. A similar conclusion is reached for amino acids and proteins.\par

\end{abstract}

\noindent
keywords: information orientation chirality polarity helix globule motor catalyst cyclicity

\begin{multicols}{2}

\section{Introduction}

A central concept of modern biology is that of the genetic material, the carrier of hereditary information, and important issues present in the foundations of biology can be explored through a specific investigation of this concept. It is known today that hereditary information has a material basis made of deoxyribonucleic acid or DNA. Why is DNA such as it is and not otherwise? We examine this question through a theoretical construction of the genetic material, using the language and the methods introduced in a previous article \cite{Sikorav2014foundations} in order to build the foundations of biology. A similar construction is used to better understand the structure of amino acids and proteins.\par

\section{The genetic material}

It is common to introduce the genetic material following the work of Watson and Crick describing first the structure of the DNA double helix and then its replication process. \cite{Watson1953a,Watson1953c} Here we present an approach that is constructive rather than descriptive, being both deductive and inductive, and proceeds from the replication process to the structure. We deduce from universal biological phenomena the necessary asymmetries that the genetic material should possess. This genetic material is not elemental but compound. The construction incorporates as many compatible symmetry elements as possible and, therefore, ends with an ideal, Platonic final structure, which is compared with that of DNA. From this emerges a better understanding of the necessary asymmetries and of the symmetries compatible with it. We conclude that DNA appears to be both unique and ideal.\par
The first steps of the construction apply equally well to polymeric nucleic acids as to proteins. The underlying reason is that both types of informational biopolymers, though different, share the same fundamental asymmetries. We can construct proteins, starting with their monomers the amino acids, using a similar approach, again leading to a better intuitive understanding of these polymers.\par

\subsection{A list of requirements}

Our goal is to build a device that contains genetic information. The properties that the genetic material should possess can be extracted from the four fundamental theories of life: \cite{Sikorav2014foundations}
\begin{itemize*}
\item
Both the theory of natural selection and informational theory of life require a transmission of hereditary information, a certain set of instructions. The minimal number of information bits contained in this set can be inferred from the large number of hereditary traits (or genes) always present in a living organism (estimated at about one thousand or more) and from the large number of bits contained in each of them (much larger than unity, about one hundred or more). This indicates that the set should contain at least $10^5$ bits of information.
\item
The universal existence of cells and of cell replication indicates that the transmission of genetic information is related to the physical process of cell division. This implies not only the existence, but also the replication and the segregation of a material entity through an appropriate transport process. The segregation of the genetic material during cell division could possibly rely on Brownian motion alone. However, this process is far more efficient when assisted by molecular motors. We therefore conclude that there must exist such motors, translocating the genetic material during cell division. It is, in turn, necessary that the structure of the genetic material be such as to allow it to interact most efficiently with these molecular motors.
\item
Hereditary information is stably conserved in cryptobiosis, during which metabolism is fully suspended. The permanence of the genetic material and the maintenance of information cannot be solely based on a continuous dissipation of energy and must be understood through the stability of an isolated system. Since the inclusion of symmetry elements increases the stability of conservative systems, this provides a strong argument to incorporate as many symmetry elements as possible in the structure to be built.
\item
The genetic material must be parsimonious in terms of amount of matter used and space occupied. This means that considerations of miniaturization and dense packing will constantly be present in the construction.
\item
Lastly, the presence of chiral compounds in living organisms raises an additional question as to the chirality of this material element. The genetic material could be chiral or else achiral, the chirality of living matter being transmitted in an epigenetic manner. The construction will clarify this issue, showing that molecular chirality is a necessary feature of the genetic material.
\end{itemize*}

\subsection{The construction}

We now present step-by-step a theoretical construction of the genetic material.

\subsubsection{Stability and information: a heteropolymer, both long and flexible}

Matter is made of atoms, and this leads to the building of a discrete structure made of a finite number of components. In order to increase the stability of an structure built of atoms, we must use the strongest chemical bonds, which are covalent (as also suggested by Schr{\"o}dinger \cite{Schroedinger1944}). The genetic material is thus, at first, a molecule. A small molecule contains little information; Therefore, to carry genetic information, the desired structure must be a macromolecule. By simplicity we shall consider a single, linear, unbranched polymer containing all the hereditary information, corresponding to the case of a unique linkage group.\par
A polymer made of identical monomers (a homopolymer) does not contain information. We thus need a heteropolymer made of at least two types of monomers. A simple binary code results from the use of two types of monomers, a higher order code if the number of distinct monomers is greater. The sequence of the monomers encodes the genetic information, which is of digital nature. This discrete structure of the monomers reflects the principle of atomicity and is compatible with the necessity of a random, quantum mechanical process involving particulate changes in the sequence as the basis of mutations. The polymeric character of the genetic material may be viewed as an expression of the principle of continuity, through the covalent bonds connecting the monomers, a molecular transposition of the concept of a linkage group. Yet, the universal process of genetic recombination in which the order of genes in a linkage group can altered implies that the continuity of the chain can be transiently interrupted. This points to the necessary existence of cut and paste devices (catalysts and motors) operating on the heteropolymer to move a portion of the chain from one place to another.\par
Assuming that each monomer contains one bit of information leads to envision a polymeric chain with a large number of monomers ($\sim 10^5$). Furthermore, as each monomer is minimally of atomic size which yields a contour length of at least 10~micrometers for this polymer. An extended conformation of this chain is incompatible with the size of a small cell or single-cell prokaryotic organism (less than one micrometer) which implies that the polymer must be flexible. Flexibility is also associated with the possibility of filling space in a parsimonious manner, thus adopting a dense, globular conformation (further described below).\par
The polymeric nature of the genetic material implies that heredity must also be understood in the language of polymer science, using the concepts of polymer physics: static conformations as well as chain dynamics. The very large degree of polymerization of the genetic material, in particular, has consequences that can be approached through scaling concepts reflecting the symmetry of scale invariance. \cite{deGennes1979}\par

\subsubsection{Fine structure of the monomers: helicity, homochirality and isotacticity}

The heteropolymer carrying the genetic information must be transported efficiently to the two daughter cells during cell division. Brownian motion alone is not be sufficient for that goal, as it can be stalled be the presence of obstacles. The requirement of transport leads to assert the existence of molecular motors, called translocases, able to translocate along the chain (or to translocate the chain if the motor remains immobile).\par
Any arbitrary displacement in space can be decomposed into a combination of a rotation around an axis and a translation. The displacement of the motor relative to the heteropolymer will be most efficient if it occurs in a regular manner, through a repetition of identical steps of translation and rotation. The polymer, as seen by the motor, must be able to adopt, at least transiently, a regular structure. As a general rule, the construction of an object permitting such regular steps leads naturally to a circular helix. \cite{Coxeter1961} A regular, chiral helix is thus the most general structure (an achiral, degenerate helix being a singular case).\par
The requirement of interaction of a motor with a regular structure implies that each monomer must contain an identical chemical group used for the interaction with the motor. We call this group a vertebra. To the vertebra is attached one of two possible side-chains, specific for the two monomers, separated from the vertebra by a chemical group used as a spacer so as not to interfere with the action of the translocase. The existence of this spacer is required to offer a regular, periodic landscape to the translocases. The vertebra is used to connect the monomers between themselves to form the polymeric chain; This chemical group is therefore trifunctional.\par
The chain of vertebrae is called the backbone of the heteropolymer. The motor must also be able to move in a constant direction along the backbone, thus to dissipate energy anisotropically when it interacts with a vertebra which must have a polar (directional) structure. To the polarity of the vertebrae is associated a polarity of the backbone, and, thus, a polarity of the genetic information encoded by the sequence of the monomers. This polarity appears as an expression of the principle of fine division: the mechanical efficiency of the translocase is based on a fine, one-dimensional orientation of the heteropolymer.\par
Each monomer thus contains a trifunctional, chiral vertebra: two of the functions are present in a polar, oriented backbone and the third connects this backbone to specific achiral side chains or residues, held outside of the reach of the motor.\par
The trifunctional vertebra need not be chiral but only prochiral. However, the simplest trifunctional vertebra contains a carbon atom with a double bond, and the resulting planar chemical structure is less flexible than a carbon atom having four sigma bounds. As we require the chain to be flexible, we are thus led to choose as a minimal vertebra a compound with a general formula HC$X_1X_2X_3$, where the three $X_i$ groups differ from one another and also from the hydrogen atom. One group contains the spacer connecting to the specific side chains and the two remaining groups are used for the polymerization of the backbone. Each vertebra is, therefore, chiral. The monomers are homochiral compounds and the polymers are isotactic chains. We can again understand this result as a consequence of the principle of fine division: a complete, three-dimensional orientation of the vertebra contributes to the mechanical efficiency of the motors through a narrow channeling of the dissipation of the chemical energy.\par
We can further assume, by simplicity, that the side groups are achiral in the absence of other specific requirements. Additional arguments in favor of the chiral character of the heteropolymer come from considerations of dense packing (not detailed in the present work).\par

\subsubsection{The replication process and its consequences: semi-conservative duplication and double helical structure}

The process of duplication of the genetic material must minimize space and maximize efficiency and economy. It must then be a local process, relying on the concept of molecular complementary recognition, following the ideas presented by Friedrich-Freksa \cite{Friedrich-Freksa1940} and by Pauling and Delbr{\"u}ck. \cite{Pauling1940,Muller1947} This concept can be seen as yet another illustration of the principle of fine division (as shown for biological catalysis by Fischer, Pauling and others). As a general rule, a mechanical copy of an object based on a molding process does not lead to a copy of the object itself, but to an object with a complementary structure. The only exception of this rule occurs if the object to be copied contains its own complement. \cite{Pauling1948} In this case, the replication process is semi-conservative. Starting from a parental structure to be replicated made of two complementary parts, one obtains two copies, each of which contains one of the two parental parts. Each complementary part of the parental structure has been used as a template for the production of the other part. The parental complementary parts end up separated at the completion of the process. Thus, replication is a duplication process, leading to an efficient exponential amplification of the initial structure as a function of the number of replication rounds. We thus conclude that in order to be copied efficiently, the genetic material must contain its own complement and is replicated through a semi-conservative process. The helical structure built so far is only half the desired structure. The complete structure is to be made of two complementary helical molecules (heteropolymers) with identical backbones.\par

\subsubsection{Fine structure of the double helix: complementary side groups held by non-covalent interactions}

During replication, each of the two chains act as template for the copy of the complementary strand. The complementariness between the two helices results from a complementariness of the specific side groups of the monomers. Assuming the existence of only two side groups in one helix forming a sequence, and of two corresponding complementary side groups, leads to the employment of a total of four different side groups, except for the particular situation where the two side groups of one helix are complementary to each other, in which case only two types of side groups would be sufficient. As this particular case is the simplest choice, we shall retain it in the following with no loss of generality.\par
Due to the complementariness rule, the two complementary monomers are to be found in equal amount in the double helical structure. In this double helix, the information is doubly present because of the complementary rules. We have seen that this redundancy can be understood not only in terms of a greater efficiency of the replication process (exponential amplification), but it is also a requirement due to the necessary imperfection of the replication process in order to fulfill the conditions of Shannon's theorem, as explained previously. Redundancy contributes to increase the stability of the genetic information in replication and in conservation.\par
The interactions between the complementary parts must be strongly attractive, yet the individual bonds are broken during the duplication process, in contrast with the covalent bonds that connect the monomers insuring the fidelity of each strand. This points to a complementariness between the side groups which relies exclusively on weak, non-covalent bonds. The two complementary strands are thus specifically held together through multiple weak (non covalent) bonds. The general role played by multiple weak attachments in the establishment of specificity had been foreseen by Pauling \cite{Pauling1948} and, more explicitly, by Crane. \cite{Crane1950} The resulting double helix is a supramolecular entity. The semi-conservative duplication process, with its need for a physical separation of the complementary parts (each of them ending up in a different daughter cell), also points to an obvious role for one of the translocases postulated to interact with the structure: to act as a strand helicase, assisting the mechanical disruption of the double helix by cleaving the weak bonds between the complementary parts.\par

\subsubsection{Temporal ordering along each strand and between strands}

The semi-conservative duplication process creates a temporal asymmetry between the two strands of the double helix: one older, that we call the $C$ strand, coming from the parental molecule, and one newly synthesized complementary strand, the $W$ strand. This asymmetry makes possible the labeling of either the parental or the daughter strand through certain chemical modifications (such as methylation). In addition, the synthesis of a complementary strand is a templated, out-of-equilibrium polymerization. This creates a temporal order of the monomers within each strand. Both temporal orderings, along each strand and between the two strands, make the structure fully oriented in time as expected from the principle of fine division.\par

\subsubsection{Geometry of the double helix: plectonemicity and antiparallel strands}

Because of the presence of complementary groups connecting the two backbones, the two polymer chains cannot be far apart from each other. Furthermore, the search for an economy of space also leads to require that the two polymers be held in close vicinity. The simplest such structure that can be envisioned of is a planar ladder, where the complementary side groups provide the rungs, but this degenerate helix is incompatible with the chiral nature of the monomers and can be rejected as singular.\
par
A simple thought experiment of the duplication of a chiral helical mathematical curve consists in copying and moving the curve by a small translation along the helical axis or rotation about this axis (meaning that the displacement is small with respect to the pitch of the helical curve). This is equivalent to the winding of two parallel lines on a cylinder and creates a pair of helices. A double helix made of two such identical chiral helices can be one of two types: a pair of helices side-by-side, called paranemic (\foreignlanguage{greek}{par\'a} 
meaning side by side, and \foreignlanguage{greek}{n{\~h}ma} 
meaning thread), or plectonemic (\foreignlanguage{greek}{plekt{\'o}c} 
meaning intertwined). In the case of a plectonemic structure, it is possible to embed the two helices within a single cylinder in a regular manner, confounding their axes of symmetry, without deformation. This is impossible to do with a paranemic structure which is, therefore, irregular and is to be ruled out. We are left with a plectonemic double helix, also expected to be more stable and more compatible with efficient interactions with molecular motors.\par
Given the polarity of the backbone, the two strands of the double helix can have either a parallel $\uparrow\uparrow$ or antiparallel $\uparrow\downarrow$ orientation. We can increase the symmetry of the structure by relating the two backbones through an appropriate geometrical transformation, namely an isometry. A reflection is ruled out by the presence of chiral elements in the backbone which means that a parallel orientation of the two backbones does not add any new symmetry elements. The symmetry can be increased through the introduction of two-fold ($C_2$) rotation axes if the two strands are coupled with opposite polarities. We therefore retain an antiparallel structure.

\subsubsection{A cyclic chain}

The double helical structure could be linear, but the presence of ends of the double helix would destroy many of the symmetries introduced above (helical symmetry and two-fold rotation axes). Indeed, there are no exact symmetries in a finite, linear helical structure. In a linear, double-helix made of a finite number of monomers, there is, in fact, only one nearly-exact $C_2$ rotation axis in the structure (this central rotation axis is only exact if we ignore the temporal asymmetry between old and new strands).\par
We can increase the symmetry of this structure by requiring it to be cyclic, eliminating the asymmetries related to the ends. In doing so, we have to bend the helix and, therefore, formally to destroy the exact symmetries associated with the helical axis. However, as the polymer is very long and flexible, both the deformation and its the energetic cost should be very small. We therefore choose to circularize the structure resulting in a covalently-closed, cyclic double-helix. The helical symmetry and the two-fold axes are not exact, but approximate, plesiosymmetries.\par

\subsubsection{A strand passing catalytic activity is required to separate the complementary strands}

An objection can be raised against the cyclic structure. Indeed, the two complementary strands forming the double helical structure are separated to become fully segregated in the daughter cells. The plectonemic structure of the cyclic double helix creates a topological obstacle to this separation, as the two complementary strands are initially catenated. This problem can be solved by appropriate catalysts performing strand-passage reactions, that we call dianemases, operating through cycles of controlled breakage and reunion. The existence of similar cut-and-paste catalysts is, in fact, already required in order to explain the process of genetic recombination. The simplest dianemase catalyzes the passage of one strand through the other by the transient breakage of the backbone of one strand followed by its resealing after the passage of the other strand through the transient breach. In the presence of this strand passing activity, an insoluble topological problem is replaced by a soluble rheological one: the two closed complementary strands can now flow one through another on a finite time scale in a process which can be assisted by molecular motors such as the helicases described above. The most efficient manner for this catalyst to operate is to become transiently covalently bound to the broken strand.\par

\subsubsection{Scaling laws for the double helix: Economy of space and globularity}

We have now obtained a heteropolymer having a double helical structure. To a first approximation, the complementary paired chains can be viewed as a homopolymer, locally difficult to bend to a helical structure (and thus semi-flexible), but globally flexible in view of its large degree of polymerization. Modern polymer physics has taught that the static and dynamic properties of long flexible homopolymers have a universal character independent of the molecular details that can be expressed in terms of scaling laws reflecting a symmetry of scale invariance. \cite{deGennes1979}\par
Given the large degree of polymerization of the double helical structure, it is reasonable to attempt a scaling analysis of its packing law. A flexible polymer can exist in one of two extreme conformations: either fully stretched or densely packed. Other, intermediate conformations are possible, such as the swollen coil and the ideal, Gaussian or Brownian conformation. These four conformations can be described through scaling laws, universal relations between the degree of polymerization $N$ and the volume $V$ occupied by the polymer. The characteristic size of the polymer is given by $R(N) \propto N^{\nu}$, where $\nu$ is called the swelling exponent (its reciprocal is called the fractal or Hausdorff dimension of the chain \cite{Grosberg1997}). The value of the scaling exponent is $\nu = 1$ for a stretched chain, about $3/5$ for a swollen coil, $1/2$ for a Brownian chain and $1/3$ for a dense, globular state. In other words, a search for a dense occupation of space within the cell implies the volume $V(N) \propto R^3$ must scale linearly with the chain length. The scaling law applies to the chain taken as a whole; other regimes could apply at lower scales. This overall globular state is expected to be permanent in a cell (as opposed to the more transient double helical conformation, which only persists between two rounds of replication). There can exist, however, a great variety of such globular states and their density can vary throughout a cell gemination cycle or with different physiological conditions. For instance, we expect the density of the polymeric globule to be higher in a dehydrated dormant cell than in a similar cell that is metabolically active. At high enough densities, the structure of this globular semi-flexible chain is expected to be locally anisotropic due to excluded volume effects between chain segments \cite{Onsager1949,Khokhlov1981} and may exhibit a liquid crystalline (nematic) ordering. The chiral structure of the double helix will favor a twisted nematic (cholesteric) ordering if the packing density is not too high.\par

\subsection{Summary of the construction}

The final, transient, structure obtained is a semi-flexible, compact (globular), cyclic, two-stranded structure. The two cyclic strands are topologically linked. The separation of the two strands occurring during the replication process is assisted by motors (called helicases), disrupting the non-covalent bonds between them, and by enzymes performing strand-passage reactions through cut-and-paste operations which contribute to decrease their topological linking number to zero. We call these enzymes dianemases, following the recommended nomenclature for naming biological catalysts (those acting on DNA have been called topoisomerases).\par
The genetic material is an information carrying device fully oriented in time and in space (as shown by the construction and as expected from the principle of fine division). The structure is characterized by necessary asymmetries: it contains information and this information is encoded both spatially and temporally: The encoding is found in the sequence of monomers and in the fact that each strand is a fully oriented polymer, both spatially (through polarity and chirality) and temporally, as there exists a temporal order along each strand, albeit imperfect, the monomers being assembled by a directed, out-of-equilibrium polymerization. The temporal ordering is also found at the level of the double helix, where one strand, which we call the $C$ strand, used as a template in the previous round of duplication, is older than the complementary, younger $W$ strand. This last asymmetry is a consequence of the necessary semi-conservative nature of the replication process. The structure obtained is ideal, having been systematically saturated with compatible symmetries: helical symmetry, homochirality, isotacticity, plectonemicity, two-fold rotation axes, circularity, globularity, complementarity, parity and redundancy, achiral side groups. None of these symmetries are exact, but only nearly so. The efficiency of the interaction with motors results both from asymmetries (polarity and chirality) and symmetries (helical: homochirality, isotacticity; two-fold rotation axes).\par
Temporal asymmetries signal the historicity of the double helix. The asymmetry between the old and the new strand makes possible their labeling through certain chemical modifications, called epigenetic, used for example in error-correction processes. In a general manner, epigenetic phenomena result from necessary asymmetries, present not only in the genetic material but also in its environment (for instance in cell membranes, where the existence of new poles following cell division can be exploited similarly).\par
We have attempted to provide the simplest constructive approach based on our current understanding of biology, making use of ideas that were unknown in 1953. This is the case for the concept of molecular motors (such as RNA polymerase \cite{Yin1995}) acting on the genetic material which only emerged in the mid 1990s, and is crucial to our approach. Similarly, the reasoning by which one can reject a paranemic (side by side) double helical structure as being irregular \cite{Watson1953b} requires to be fully rigorous C\u{a}lug\u{a}reanu formula relating the linking number of closed curves to twist and writhe. \cite{Calugareanu1961,White1969,Fuller1971}

\subsection{Comparison with the structure of nucleotides and nucleic acids}

\begin{figure}[!ht]
\begin{center}
\includegraphics{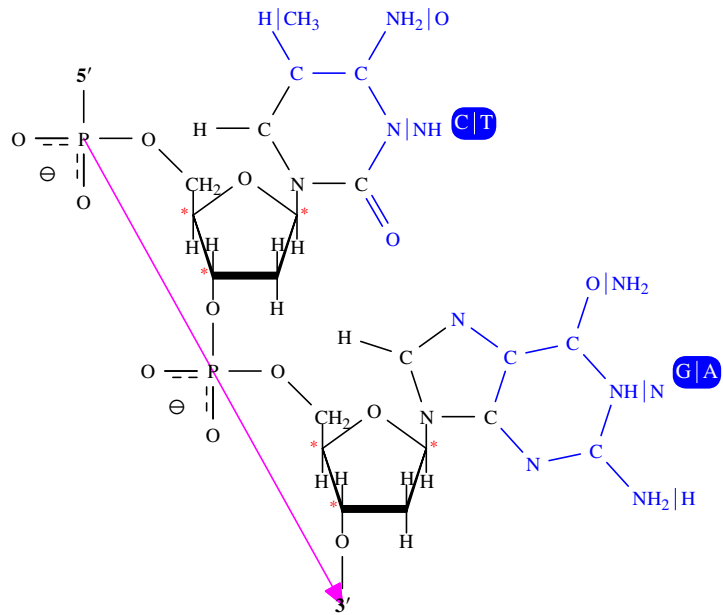}
\end{center}
\caption{{\bf Structure of nucleotides and primary structure of polynucleotides. I Deoxyribonucleic acid.}
Two consecutive monomers are shown. The planar nucleobases (C: cytosine, T: thymine, G: guanine, and A: adenine) include atoms drawn in black belonging to the backbone, being common to all of the bases; the atoms that differ are drawn in blue and constitute the specific residues. The magenta arrow indicates the polarity of the backbone. The three chiral carbon atoms are indicated by red asterisks.}
\label{fig:DNA}
\end{figure}

\begin{figure}[!ht]
\begin{center}
\includegraphics{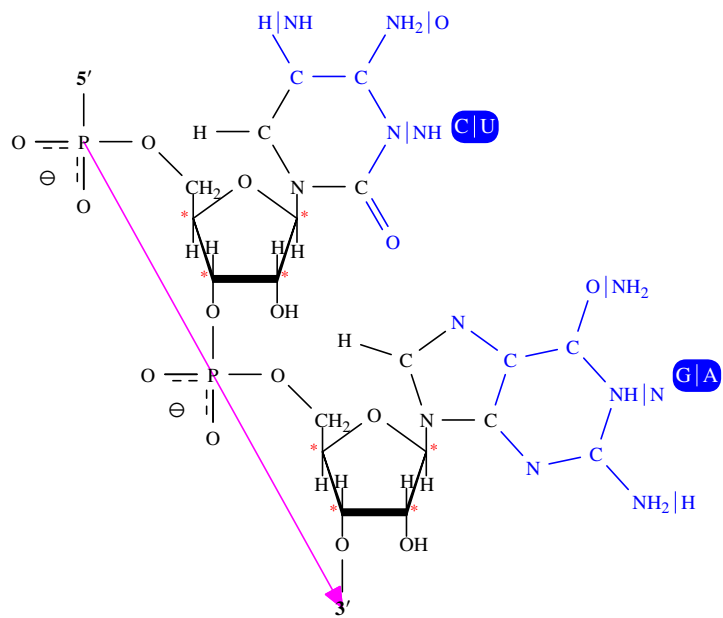}
\end{center}
\caption{{\bf Structure of nucleotides and primary structure of polynucleotides. II Ribonucleic acid.}
The same convention is used as in Figure~\ref{fig:DNA}, with U designating uracil. The backbone contains a fourth chiral carbon atom and a reactive hydroxyl group.}
\label{fig:RNA}
\end{figure}

The theoretical construction allows one to better understand the structure of nucleic acids; we focus here on the primary structure of DNA and RNA, shown in Figures~\ref{fig:DNA} and \ref{fig:RNA}. DNA and RNA are heteropolymers made of four monomers (rather than the predicted, minimal number of two). The monomers, called nucleotides, consist of a trifunctional handle to which is attached one out of five planar, achiral nucleobases. The molecular vertebra contains a trifunctional deoxyribose or ribose, to which are attached the $5^{\prime}$ and $3^{\prime}$ phosphates: these two links account for the polarities of the vertebra and of the heteropolymeric chains. The vertebra is both polar and chiral as expected, but its structure is not minimal (that is consisting of a single stereogenic carbon): instead, the three stereogenic carbons located in the pentose ring are endowed each with one of the three specific functions of the vertebra, being associated to the $3^{\prime}$ end (for $\textrm{C}_3^{\prime}$), to the $5^{\prime}$ end (for $\textrm{C}_4^{\prime}$), or to the lateral base ($\textrm{C}_1^{\prime}$), as described by Natta. \cite{Natta1968} The planar side groups are always connected in the same manner, forming an isotactic chain. RNA contains a fourth chiral $\mathrm{C}_2^{\prime}$, lowering its symmetry, thus decreasing its stability. This additional asymmetry is associated with novel phenomena: the attached hydroxyl group is a catalytic component of several ribozymes. The phosphorus atom as well as all non-chiral carbon atoms are prochiral (with the exception of the methyl group of the thymine base of DNA). All atoms attached to tetravalent phosphorus and carbon atoms are thus discernible. Similarly, the two faces defined by (trigonal) trivalent carbons and their attached groups are distinct, illustrating again the principle of fine division. Lastly, both in DNA and RNA, the common vertebra extends, in fact, into the planar nucleobases. Indeed, four atoms are common in all bases, not only the nitrogen linked to the deoxyribose or ribose, but also two carbon atoms connected to this nitrogen and an additional hydrogen atom. (To emphasize this fact, we have drawn the structure in Figures~\ref{fig:DNA} and~\ref{fig:RNA} using different colors for the atoms belonging to the backbone, common to all monomers, shown in black, and for the atoms of the bases specific to each residue, colored in blue.) The geometry of the common atoms associated with the backbone only differ minimally by the angles of six- or five-atom heterocyclic ring structures of the purines and pyrimidines ($60^{\circ}$ versus $72^{\circ}$). The differences of chemical structure between the four nucleobases arise by specific substitutions of reactive groups in the pyrimidines cytosine, thymine (and uracil) as well as in the purines guanine and adenine.\par
The minimal number of monomers required to synthesize informational polymers is two. The existence of four bases, thus of two distinct types of complementary base pairs having different bond strengths (involving two hydrogen bonds for adenine with thymine versus three for guanine with cytosine), can be explained as follows: In the helix-coil denaturation of DNA, the secondary structure of the double helix shows a sequence-dependent stability, a phenomenon of central importance in replication, transcription and recombination. \cite{Yeramian2000} This would be impossible to observe in a double helical polymer made of only two types of monomers, thus of a single type of pair. Indeed, the stability of such a double helix would be that of a homopolymer. The additional asymmetry makes possible the orientation of the double helical structure of DNA as well as novel phenomena in the tertiary structure of chromosomes, for example, or in the interaction with catalysts.\par
An unreplicated chromosomal DNA molecule consists of a single double-helical polymer (this is also known as the mononeme hypothesis). The overall conformation of this extremely long polymer is indeed globular within cells. This general statement refers to a broad and complex field of investigation, that of DNA condensation to be discussed elsewhere. In contrast with globularity, circularity seems at first not to be universally observed. Circular chromosomal DNA appears widespread in prokaryotes and is probably universal. Eukaryotic chromosomes are usually linear (circularized versions of these chromosomes can also be observed in mutants, usually associated with genetic deficiencies). Furthermore, eukaroytic cells always contain cyclic DNA molecules as episomes (such as in mitochondrial DNA). We formulate the hypothesis that all cells contain a cyclic DNA molecule, either as a chromosomal chain or as an episome. Circular DNA and the process of DNA cyclization appears to be universal among living organisms.\par

\section{The structure of amino acids and proteins}

Proteins are heteropolymers that are assembled by a molecular motor called the ribosome which must translocate efficiently during the polymerization steps. This implies that one can apply the analysis leading to the conclusions reached above concerning the genetic material to understand the structure of amino acids and proteins.
\begin{enumerate*}
\item Proteins must be assembled from monomers containing trifunctional vertebrae that are polar, chiral and to which side groups are attached;
\item The side groups should be achiral, spatially separated by a spacer from the molecular vertebra (to avoid direct contact with the motor).
\item Polymers assembled from these monomers, polypeptides and proteins, must also be able to adopt, at least transiently, a helical conformation.
\item The tertiary structure of these polymers should be globular to minimize space occupation.
\end{enumerate*}
\par
\begin{figure}[!ht]
\begin{center}
\includegraphics{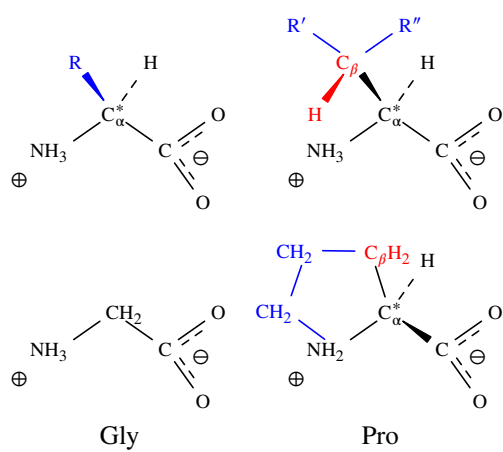}
\end{center}
\caption{{\bf Chemical structure of proteinogenic amino acids.}
Top, left: conventional generic representation of the amino acids showing a residue $\mathrm{R}$ (in blue) attached to the asymmetric $\mathrm{C}_{\alpha}^{\ast}$ atom;
Top, right: refined generic representation, emphasizing the presence of a prochiral spacer group $\mathrm{C}_{\beta}H$ (in red) to which are attached two residues $\mathrm{R}^{\prime}$ and $\mathrm{R}^{\prime\prime}$ (in blue). See also Table~\ref{tab:aminoacids}.
Bottom, left: glycine is achiral and lacks a methyne spacer group.
Bottom, right: proline is also somewhat an anomaly since its side group cycles back to the amino group of the vertebra.
}
\label{fig:aminoacids}
\end{figure}

The structure of the twenty proteinogenic amino acids \cite{Crick1958} obeys these basic rules. The standard representation of these amino acids and of the polypeptidic chain is due to Fischer. It is based on the concept of the side group, denoted $\mathrm{R}$, specific for each amino acid and attached to the asymmetric $\mathrm{C}_{\alpha}^{\ast}$ atom, as illustrated for the monomer in Figure~\ref{fig:aminoacids} (top, left) and written below for a two amino acid peptidic chain: \cite{Fischer1904}
$$ \mathrm{NH}_2.\mathrm{C}^{\ast}\mathrm{HR}.\mathrm{CO}.\mathrm{NH}.\mathrm{C}^{\ast}\mathrm{HR}.\mathrm{COOH} $$
We propose a refined representation of these amino acids, shown in Figure~\ref{fig:aminoacids} (top, right) which emphasizes the existence of a $\mathrm{C}_{\beta}\mathrm{H}$ spacer group. The structure minimizes the amount of matter required to build a vertebra: the three-point, chiral, handle consists of a single chiral carbon to which is attached a hydrogen atom and the constitution of the spacer consists of a single methyne group. These structures cannot be further reduced. One can describe this simplicity in terms of atom economy or biological perfection. This economy results from the high energetic cost of the synthesis of proteins (the cost of nucleic acid synthesis is in comparison much lower, and appears compatible with their more opulent sugar-phosphate-aromatic vertebra).\par
The molecular vertebra consists of a single, trifunctional asymmetric $\mathrm{C}_{\alpha}^{\ast}$ atom to which are also attached hydrogen, carboxyl and amino reactive groups. The description of the molecular vertebra now includes a methyne spacer, which can be removed from the conventional specific residues or side groups. The $\mathrm{C}_{\beta}$ atom of this spacer is predicted and observed to be achiral (isoleucine and threonine are exceptions to this prediction, and the absolute configuration is S for Ile and R for Thr in the Cahn-Ingold-Prelog notation). Table~\ref{tab:aminoacids}, listing the twenty proteinogenic amino acids sorted by decreasing molecular weight, details the simpler residues that consist of two groups $\mathrm{R}^{\prime}$ and $\mathrm{R}^{\prime\prime}$. The second residue $\mathrm{R}^{\prime\prime}$ is a single hydrogen atom except for isoleucine, threonine and valine (where $\mathrm{R}^{\prime\prime}=\mathrm{CH}_3$). Glycine is special as it is achiral and lacks a methyne spacer group. However, the side group of glycine, being a simple H atom, causes no steric clash with a molecular motor and glycine often plays the role of a flexible junction in the construction of protein chains. Note that the carbon atom of glycine is prochiral and that the two hydrogen atoms attached to it are discernible.\par
Table~\ref{tab:aminoacids} also lists the molecules obtained when a hydrogen atom would be added to the separated active residue $\mathrm{R}^{\prime}$ in place of the spacer bond (in the case of proline, where the side group is cyclically attached to the backbone, a second terminating hydrogen atom is included). This permits to identify sixteen trimmed specific active side groups. The functional properties of the trimmed side groups are more readily grasped in this new representation than when the $\mathrm{C}_{\beta}\mathrm{H}$ spacer is included in the residue: the specific side group $\mathrm{H-R}^{\prime}$ of tryptophane is thus indole (instead of skatol), that of tyrosine is phenol, that of phenylanine is benzene, and that of histidine is imidazole. The structure of these amino acids appears minimal if one wants to employ the corresponding functional groups. In contrast, this is not the case for other amino acids: for instance, lysine with 1-propylamine or arginine with N-ethylguanidine, where the functional group (amine or guanidine) could be attached using a short spacer. Altogether, eleven out of the twenty amino acids appear to possess minimal side groups (Trp, Tyr, Phe, His, Met, Asp, Asn, Cys, Ala and Gly).\par
The novel representation that we present here does not aid our understanding in the case of short side chains (alanine, cysteine and serine), as the removal of the methylene group leaves the residue without a carbon atom. Here, it is be better to associate the spacer with the functional side group, leading to the molecules methane, methylmercaptan and methanol (instead of dihydrogen, hydrogen sulfide and water).\par
The major role of proteins as catalysts implies, in contrast to the relative simplicity of the genetic material, that proteins should contain the main functional groups of organic chemistry, such as acid and base, alcohol and thiol, or aromatic (benzene, indol and phenol). Proteins, furthermore, are to be produced in greater quantities then the genetic material itself and, therefore, must involve an economical use of matter (atoms).\par
The conclusion reached above that eleven out of the twenty proteinogenic amino acids possess minimal side groups implies that the necessary number of functional groups is greater than the number of strict minimal structures.\par
The analysis developed here permits to understand the polar (in the sense of a temporal and spatial, one-dimensional, orientation) and chiral structure of the backbone of proteins and their ability to adopt, at least transiently (during the elongation of the polypeptidic chain by the ribosome machinery), a helical configuration. In particular, the observation that the conformation of amino acids in polypeptides and proteins is very similar in Ramachandran plots \cite{Ramachandran1963,Hovmoeller2002} for all amino acids (except for the two exceptions glycine and proline, which are associated with a disruption of helical order) can be rationalized in terms of the requirement of an efficient, side chain independent, interaction with a molecular motor.\par
Many proteins, in particular enzymes, are folded in dense, globular structures in their native, functional state. As in DNA, densely packed structures are to be expected as a consequence of the constant search for miniaturization by natural selection. Also as for DNA, this overall dense conformation can be described by a universal scaling law relating the volume of a globular protein and its degree of polymerization. The fractal exponent determined experimentally for proteins ($1/v = 2.6$) is, in fact, slightly smaller than that of a completely compact, three-dimensional collapsed polymer, \cite{Enright2005} a result which can be viewed as a consequence of the small number of monomers that make enzymes in comparison with the much larger number of monomers present in chromosomal DNA.\par
Lastly we observe that proteins are usually linear. However, as for DNA, it is tempting to speculate that the existence of cyclic proteins is universal among living organisms. In support of this hypothesis, it can be observed that a growing number of cyclic proteins has been recently described. \cite{Trabi2002}\par
These investigations suggest that it will be eventually possible to explain why the proteinogenic amino acids, their common vertebra and their specific side groups are such as they are and not otherwise.\par

\section{Comparison of nucleic acids and proteins}

Both nucleic acid chains (DNA and RNA) and proteins are heteropolymers assembled out of equilibrium by molecular motors. They share common invariants, common asymmetries (presence of information, spatial orientation through polarity and chirality as well as temporal orientation) and common symmetries (helical symmetry, homochirality, isotacticity, globularity) and achiral side groups (except for two amino acids out of twenty). These common invariants possess the antiquity of life itself. However, they fulfill mostly different functions: nucleic acids forming the genetic material provide a memory having long-term stability and reliability and proteins hold structural roles or serve as biochemical catalysts.\par
The central tools used in the theoretical constructions, not only catalysts but also biological motors, must have been already present at a very early stage, raising a general question of the prebiotic evolution of such motors.\par
No more than four monomers are required in the constitution of nucleic acids, whereas at least eleven monomers are necessary in proteins. These functional as well as structural differences appear to reflect a certain division of labor between nucleic acids and proteins, the genetic material and catalysts. This suggests a plausible explanation for the necessary existence of the two types of biopolymers. It also points to the likely uniqueness of the logical solution offered by von~Neumann for self-reproduction.\par
The results of the constructions described above constrain the structures of biopolymers and their constituent monomers. This can contribute to the design of non-natural nucleotides or amino acids to be used in synthetic biology, to a better understanding of prebiotic chemistry, and in the search for extraterrestrial forms of life.\par

\paragraph*{Acknowledgements}
The present work summarizes an ongoing research program initiated a few years ago. Previous versions of it have been presented to various audiences: Carg{\`e}se summer school on DNA and Chromosomes, in 2004, 2006, and 2009; Gent Fantom Research School on Symmetries and symmetry violation in 2004 and 2007; Lectures on the Foundations of Molecular Biology, Evry in 2004; Lectures on Elements of Biology, Ecole Polytechnique F{\'e}d{\'e}rale de Lausanne in 2008; Lectures on Foundations of Biology, University Pierre and Marie Curie, Paris 2008-2013; 74th Cold Spring Harbor Symposium ``Evolution --- The Molecular Landscape'' in 2009 and Meeting ``From Base Pair to Body Plan'', Cold Spring Harbor Laboratory in 2013. We have greatly benefited from the comments of their participants. We wish to thank for discussions and/or comments the manuscript Roger Balian, Sydney Brenner, George Church, Gregory Chaitin, Albert Libchaber, Marie-Claude Marsolier-Kergoat, Theo Odijk, Monica Olvera de la Cruz and Edouard Y{\'e}ramian.\par

\end{multicols}


\begin{thebibliography}{10}

\bibitem{Sikorav2014foundations}
Sikorav JL, Braslau A, Goldar A.
\newblock Foundations of Biology; 2014.
\newblock Submitted to the Journal of the Royal Society Interface.

\bibitem{Watson1953a}
Watson JD, Crick FHC.
\newblock Molecular structure of nucleic acids --- A structure for deoxyribose
  nucleic acid.
\newblock Nature (London, U K). 1953;171(4356):737--738.
\newblock \href {http://dx.doi.org/10.1038/171737a0} {doi:10.1038/171737a0}.

\bibitem{Watson1953c}
Watson JD, Crick FHC.
\newblock Genetical implications of the structure of deoxyribonucleic acid.
\newblock Nature (London, U K). 1953;171(4361):964--967.
\newblock \href {http://dx.doi.org/10.1038/171964b0} {doi:10.1038/171964b0}.

\bibitem{Schroedinger1944}
Schr{\"o}dinger E.
\newblock What is Life? The Physical Aspect of the Living Cell.
\newblock Cambridge: Cambridge Univ. Press; 1944.

\bibitem{deGennes1979}
{de Gennes} PG.
\newblock Scaling concepts in polymer physics.
\newblock Ithaca: Cornell Univ. Press; 1979.

\bibitem{Coxeter1961}
Coxeter HSM.
\newblock Introduction to Geometry.
\newblock New York: Wiley; 1961.

\bibitem{Friedrich-Freksa1940}
Friedrich-Freksa H.
\newblock Bei der chromosomenkonjugation wirksame kr{\"a}fte und ihre bedeutung
  f{\"u}r identische verdopplung von nucleoproteinen.
\newblock Naturewissenchaften. 1940;28(24):376--379.
\newblock \href {http://dx.doi.org/10.1007/bf01480270}
  {doi:10.1007/bf01480270}.

\bibitem{Pauling1940}
Pauling L, Delbr{\"u}ck M.
\newblock The nature of the intermolecular forces operative in biological
  processes.
\newblock Science (Washington, DC, U S). 1940;92(2378):77--79.
\newblock \href {http://dx.doi.org/10.1126/science.92.2378.77}
  {doi:10.1126/science.92.2378.77}.

\bibitem{Muller1947}
Muller HJ.
\newblock Pilgrim Trust Lecture: The gene.
\newblock Proceedings of the Royal Society of London Series B-Biological
  Sciences. 1947;134(874):1--37.
\newblock \href {http://dx.doi.org/10.1098/rspb.1947.0001}
  {doi:10.1098/rspb.1947.0001}.

\bibitem{Pauling1948}
Pauling L. Molecular architecture and the processes of life. 21st Sir Jesse
  Boot Foundation Lecture. Nottingham, UK: The University of Nottingham; 1948.

\bibitem{Crane1950}
Crane HR.
\newblock Principles and problems of biological growth.
\newblock Scientific Monthly. 1950;70:376--389.
\newblock \href {http://dx.doi.org/10.2307/20180} {doi:10.2307/20180}.

\bibitem{Grosberg1997}
Grosberg AY, Khokhlov AR.
\newblock Giant Molecules. Here, There, and Everywhere.
\newblock San Diego: Academic Press; 1997.

\bibitem{Onsager1949}
Onsager L.
\newblock The effects of shape on the interaction of colloidal particles.
\newblock Annals of the New York Academy of Sciences. 1949;51(4):627--659.
\newblock \href {http://dx.doi.org/10.1111/j.1749-6632.1949.tb27296.x}
  {doi:10.1111/j.1749-6632.1949.tb27296.x}.

\bibitem{Khokhlov1981}
Khokhlov AR, Semenov AN.
\newblock Liquid-crystalline ordering in the solution of long persistent
  chains.
\newblock Physica A (Amsterdam, Neth). 1981;108(2--3):546--556.
\newblock \href {http://dx.doi.org/10.1016/0378-4371(81)90148-5}
  {doi:10.1016/0378-4371(81)90148-5}.

\bibitem{Yin1995}
Yin H, Wang MD, Svoboda K, Landick R, Block SM, Gelles J.
\newblock Transcription Against an Applied Force.
\newblock Science. 1995;270(5242):1653--1657.
\newblock \href {http://dx.doi.org/10.1126/science.270.5242.1653}
  {doi:10.1126/science.270.5242.1653}.

\bibitem{Watson1953b}
Watson JD, Crick FHC.
\newblock The structure of {DNA}.
\newblock Cold Spring Harbor Symposia on Quantitative Biology.
  1953;18:123--131.
\newblock \href {http://dx.doi.org/10.1101/sqb.1953.018.01.020}
  {doi:10.1101/sqb.1953.018.01.020}.

\bibitem{Calugareanu1961}
C\u{a}lug\u{a}reanu G.
\newblock Sur les classes d'isotopie des n{\oe}uds tridimensionnels et leurs
  invariants.
\newblock Czechoslovak Mathematical Journal. 1961;11(4):588--625.

\bibitem{White1969}
White JH.
\newblock Self-linking and the {G}auss integral in higher dimensions.
\newblock American Journal of Mathematics. 1969;91:693--728.
\newblock \href {http://dx.doi.org/10.2307/2373348} {doi:10.2307/2373348}.

\bibitem{Fuller1971}
Fuller FB.
\newblock The writhing number of a space curve.
\newblock Proceedings of the National Academy of Sciences of the United States
  of America. 1971;68:815--819.
\newblock \href {http://dx.doi.org/10.1073/pnas.68.4.815}
  {doi:10.1073/pnas.68.4.815}.

\bibitem{Natta1968}
Natta G, Farina M.
\newblock Stereochimica --- molecole in 3D.
\newblock Milano: Mondadori; 1968.
\newblock Stereochemistry.
\newblock London: Longman; 1972.
\newblock Translated by A. Dempster.

\bibitem{Yeramian2000}
Yeramian E.
\newblock Genes and the physics of the {DNA} double-helix.
\newblock Gene. 2000;255(2):139--150.
\newblock \href {http://dx.doi.org/10.1016/S0378-1119(00)00301-2}
  {doi:10.1016/S0378-1119(00)00301-2}.

\bibitem{Crick1958}
Crick FHC.
\newblock On Protein Synthesis.
\newblock Symposia of the Society for Experimental Biology. 1958;12:138--163.

\bibitem{Fischer1904}
Fischer E.
\newblock Synthese von Polypeptiden. {II}.
\newblock Berichte der deutschen chemischen Gesellschaft. 1904;37:2486.
\newblock \href {http://dx.doi.org/10.1002/cber.190403702197}
  {doi:10.1002/cber.190403702197}.

\bibitem{Ramachandran1963}
Ramachandran GN, Ramakrishnan C, Sasikharan V.
\newblock Stereochemistry of polypeptide chain configurations.
\newblock Journal of Molecular Biology. 1963;7(1):95--99.
\newblock \href {http://dx.doi.org/10.1016/S0022-2836(63)80023-6}
  {doi:10.1016/S0022-2836(63)80023-6}.

\bibitem{Hovmoeller2002}
Hovm{\"o}ller S, Zhou T, Ohlson T.
\newblock Conformations of amino acids in proteins.
\newblock Acta Crystallographica Section D. 2002;58(5):768--776.
\newblock \href {http://dx.doi.org/10.1107/S0907444902003359}
  {doi:10.1107/S0907444902003359}.

\bibitem{Enright2005}
Enright MB, Leitner DM.
\newblock Mass fractal dimension and the compactness of proteins.
\newblock Phys Rev E. 2005;71(1):011912.
\newblock \href {http://dx.doi.org/10.1103/PhysRevE.71.011912}
  {doi:10.1103/PhysRevE.71.011912}.

\bibitem{Trabi2002}
Trabi M, Craik DJ.
\newblock Circular proteins --- no end in sight.
\newblock Trends in Biochemical Sciences. 2002;27(3):132--138.
\newblock \href {http://dx.doi.org/10.1016/S0968-0004(02)02057-1}
  {doi:10.1016/S0968-0004(02)02057-1}.

\end{thebibliography}

\clearpage

\clearpage

\begin{landscape}
\begin{table}[!ht]
\caption{The proteinogenic amino acids and their side groups.
The table is sorted by decreasing molecular weight of the amino acids.}
\centering
\begin{tabular} {@{}|*{6}{l@{ }}p{8em}@{ }*{2}{l@{ }}p{8em}|@{}}
\hline
 & \multicolumn{3}{l@{ }}{Amino acid} & $\mathrm{M}_{\mathrm{w}}$ (Da) & $\mathrm{-R}$ (conventional) &
$\mathrm{H-R}$ & $\mathrm{-R}^{\prime\prime}$ & $\mathrm{-R}^{\prime}$ & $\mathrm{H-R}^{\prime}$ \\
\hline
1 & W & Trp & Tryptophan & 204.225 & $\mathrm{-CH}_2\mathrm{C}_8\mathrm{H}_6\mathrm{N}$ &
skatole 3-methylindole & $\mathrm{-H}$ & $\mathrm{-C}_8\mathrm{H}_6\mathrm{N}$ & indole \\
2 & Y & Tyr & Tyrosine & 181.188 & $\mathrm{-CH}_2\mathrm{C}_6\mathrm{H}_4\mathrm{OH}$ &
4-methylphenol 4-cresol & $\mathrm{-H}$ & $\mathrm{-C}_6\mathrm{H}_4\mathrm{OH}$ &
phenol \\
3 & R & Arg & Arginine & 174.201 & $\mathrm{-(CH}_2\mathrm{)}_3\mathrm{NHC(NH)NH}_2$ &
N-propylguanidine & $\mathrm{-H}$ & $\mathrm{-(CH}_2\mathrm{)}_2\mathrm{NHC(NH)NH}_2$ &
N-ethylguanidine \\
4 & F & Phe & Phenylalanine & 165.189 & $\mathrm{-CH}_2\mathrm{C}_6\mathrm{H}_5$ &
toluene & $\mathrm{-H}$ & $\mathrm{-C}_6\mathrm{H}_5$ & benzene \\
5 & H & His & Histidine & 155.155 & $\mathrm{-CH}_2\mathrm{C}_3\mathrm{H}_3\mathrm{N}_2$ &
4-methylimidazole & $\mathrm{-H}$ & $\mathrm{-C}_3\mathrm{H}_3\mathrm{N}_2$ & imidazole \\
6 & M & Met & Methionine & 149.211 & $\mathrm{-(CH}_2\mathrm{)}_2\mathrm{SCH}_3$ &
ethyl methyl sulfide & $\mathrm{-H}$ & $\mathrm{-CH}_2\mathrm{SCH}_3$ & dimethylsulfide \\
7 & E & Glu & Glutamate & 147.129 & $\mathrm{-(CH}_2\mathrm{)}_2\mathrm{COOH}$ &
propionic acid & $\mathrm{-H}$ & $\mathrm{-CH}_2\mathrm{COOH}$ & acetic acid \\
8 & K & Lys & Lysine & 146.188 & $\mathrm{-(CH}_2\mathrm{)}_4\mathrm{NH}_2$ &
1-butylamine & $\mathrm{-H}$ & $\mathrm{-(CH}_2\mathrm{)}_3\mathrm{NH}_2$ & 1-propylamine \\
9 & Q & Gln & Glutamine & 146.144 & $\mathrm{-(CH}_2\mathrm{)}_2\mathrm{CONH}_2$ &
propionamide & $\mathrm{-H}$ & $\mathrm{-CH}_2\mathrm{CONH}_2$ & acetamide \\
10 & D & Asp & Aspartate & 133.103 & $\mathrm{-CH}_2\mathrm{COOH}$ &
acetic~acid & $\mathrm{-H}$ & $\mathrm{-COOH}$ & formic~acid \\
11 & N & Asn & Asparagine & 132.118 & $\mathrm{-CH}_2\mathrm{CONH}_2$ &
acetamide & $\mathrm{-H}$ & $\mathrm{-CONH}_2$ & formamide \\
12 & L & Leu & Leucine & 131.173 & $\mathrm{-CH}_2\mathrm{CH(CH}_3\mathrm{)}_2$ &
isobutane & $\mathrm{-H}$ & $\mathrm{-CH(CH}_3\mathrm{)}_2$ & propane \\
13 & I & Ile & Isoleucine & 131.173 & $\mathrm{-CHCH}_3\mathrm{CH}_2\mathrm{CH}_3$ &
butane & $\mathrm{-CH}_3$ & $\mathrm{-CH}_2\mathrm{CH}_3$ & ethane \\
14 & C & Cys & Cysteine & 121.158 & $\mathrm{-CH}_2\mathrm{SH}$ &
methanethiol methyl~mercaptan & $\mathrm{-H}$ & $\mathrm{-HS}$ & hydrogen sulfide \\
15 & T & Thr & Threonine & 119.119 & $\mathrm{-CH(OH)CH}_3$ &
ethanol & $\mathrm{-CH}_3$ & $\mathrm{-OH}$ & water \\
16 & V & Val & Valine & 117.146 & $\mathrm{-CH(CH}_3\mathrm{)}_2$ &
propane & $\mathrm{-CH}_3$ & $\mathrm{-CH}_3$ & methane \\
17 & P & Pro & Proline & 115.130 & $\mathrm{-(CH}_2\mathrm{)}_3\mathrm{-}$ &
propane ($\mathrm{H-R-H}$) & $\mathrm{-H}$ & $\mathrm{-(CH}_2\mathrm{)}_2\mathrm{-}$ &
ethane ($\mathrm{H-R}^{\prime}\mathrm{-H}$) \\
18 & S & Ser & Serine & 105.093 & $\mathrm{-CH}_2\mathrm{OH}$ &
methanol & $\mathrm{-H}$ & $\mathrm{-OH}$ & water \\
19 & A & Ala & Alanine & 89.093 & $\mathrm{-CH}_3$ &
methane & $\mathrm{-H}$ & $\mathrm{-H}$ & dihydrogen \\
20 & G & Gly & Glycine & 75.067 & $\mathrm{-H}$ &
dihydrogen & $\varnothing$ & $\varnothing$ & $\varnothing$ \\
\hline
\end{tabular}
\label{tab:aminoacids}
\end{table}
\end{landscape}



\end{document}